\documentclass[letterpaper]{article}
\usepackage{lsrgan}
\pdfoutput=1
\usepackage{times}
\usepackage{helvet}
\usepackage{courier}
\usepackage[hyphens]{url}
\usepackage{graphicx}
\usepackage{graphicx}
\usepackage{amsmath}
\usepackage{wrapfig}
\usepackage{booktabs}
\usepackage{multirow}
\def \epsFigPath {./}
\title{Optimizing Generative Adversarial Networks for Image Super Resolution via Latent Space Regularization}
\author{Sheng Zhong\\
  Agora.io\\
  2804 Mission College Blvd, STE 110\\
  Santa Clara, CA 95054 \\
  {\tt\small shawn.zhong@agora.io}
\And
  Shifu Zhou \\
  Agora.io \\
  333 Songhu Road, Floor 8 \\
  Shanghai, China \\
  {\tt\small zhoushifu@agora.io}
}

\begin{document}
\maketitle
\begin{abstract}
  Natural images can be regarded as residing in a manifold that is embedded in a higher dimensional Euclidean space. Generative Adversarial Networks (GANs) try to learn the distribution of the real images in the manifold to generate samples that look real. But the results of existing methods still exhibit many unpleasant artifacts and distortions even for the cases where the desired ground truth target images are available for supervised learning such as in single image super resolution (SISR). We probe for ways to alleviate these problems for supervised GANs in this paper. We explicitly apply the Lipschitz Continuity Condition (LCC) to regularize the GAN. An encoding network that maps the image space to a new optimal latent space is derived from the LCC, and it is used to augment the GAN as a coupling component. The LCC is also converted to new regularization terms in the generator loss function to enforce local invariance. The GAN is optimized together with the encoding network in an attempt to make the generator converge to a more ideal and disentangled mapping that can generate samples more faithful to the target images. When the proposed models are applied to the single image super resolution problem, the results outperform the state of the art.
\end{abstract}


\section{Introduction}
A natural image can be regarded as residing in a manifold embedded in a higher dimensional space (aka the ambient space). The manifold is usually of lower dimensionality than that of the ambient space and can be mapped to a lower dimensional space by a homeomorphic function (also called an encoder in Deep Neural Network (DNN) terminology). The lower dimensional space is called the latent space. The inverse map from the latent space to the ambient space is also a homeomorphic map (i.e. the generator function). In generative DNN models such as the GAN \cite{goodfellow2014generative}, it is desired that the ideal generator function is approximated by the DNN as closely as possible.

The GAN model \cite{goodfellow2014generative} provides a powerful model to generate samples that imitate the real data. It is trained through an adversarial process involving the generator G and discriminator D. GANs suffer from problems such as mode collapse, structure distortions and training instability \cite{goodfellow2016tutorial}. DCGAN \cite{RadfordMC15} applies batch normalization to many deep layers and replaces the pooling layers with strided convolutions to alleviate the problems.
Metz et al. \cite{metz2017unrolled} define the generator objective with respect to an unrolled optimization of the discriminator to stabilize GAN training and reduce mode collapse. Arjovsky et al. \cite{arjovsky2017wasserstein,Gulrajani2017Improved} propose the Wasserstein GANs by employing the Earth Mover distance and the gradient penalty as the critic function; this helps reduce the mode collapse problem and makes the model converge more stably. Lei et al. \cite{Lei2018Geometric,Lei2017A} study generative models from computational geometry point of view, in which latent space optimization via the optimal mass transportation provides an interesting perspective. Yet the method is intractable in high dimensional spaces.
Donahue et al. \cite{donahue2017adversarial} and Dumoulin et al. \cite{dumoulin2017adversarially} propose the Bidirectional GAN (BiGAN) to extend the GAN framework to include an encoder $E: X \rightarrow Z$ that maps in the reverse direction of the generator. The BiGAN discriminator then needs to distinguish the pairs $(G(Z), Z)$ and $(X, E(X))$ with the discriminator and encoder forming another set of adversarial nets. The BiGAN often produces reconstructions of images that look little like the originals, despite often being semantically related. Rubenstein et al. \cite{rubenstein2018an} further improve the BiGAN by adding an auto-encoding loss; and they also find that simply training an autoencoder to invert the generator of a standard GAN is a viable alternative despite that the numeric quality score for the reconstructed image is inferior to BiGAN. All these help making the generated samples look more realistic or the reconstructed image more like the original. But still they are often distorted more than desired and often lack details; and internal image structures are lost in many cases.

 This is also the case when the desired ground truth target samples are available for supervised learning, as in the typical GAN application to the vision task of Single Image Super Resolution (SISR). The SISR aims at recovering the high-resolution (HR) image based on a single low-resolution (LR) image. Some noisy LR and corresponding ground truth HR image pairs are provided for supervised training. While many DNN architectures and training strategies have been used to optimize the Mean Square Error (MSE, i.e. the $L_2$-norm) or equivalently the Peak Signal-to-Noise Ratio (PSNR) \cite{ledig2017photo,lai2017deep,tai2017image,haris2018deep}, they tend to produce over-smoothed results without sufficient high-frequency details. It is found that a metric such as MSE/PSNR alone do not correlate well enough with the perception of human visual systems \cite{DBLP:journals/corr/abs-1809-07517,ledig2017photo}.

 Perceptual-based methods have therefore been proposed to optimize super-resolution DNN models with loss functions in the feature space instead of in the pixel space \cite{johnson2016perceptual,bruna2015super}. In particular, GAN is used in SISR and the SRGAN \cite{ledig2017photo} model is built with residual blocks and optimized using perceptual loss defined in feature spaces. This significantly improves the overall visual quality of reconstructed HR images over the PSNR-oriented methods.
Mechrez et al. \cite{mechrez2018learning} measured the perceptual similarity based on the cosine distance between vectors of the latent space features of the pre-trained VGG19 DNN \cite{simonyan2015very}. This helps push the generator to maintain internal statistics of images and make the output lie on the manifold of natural images and state of the art results are achieved.

\begin{figure}
  \centering
 \includegraphics[width=2.6cm,height=1.7cm]{\epsFigPath 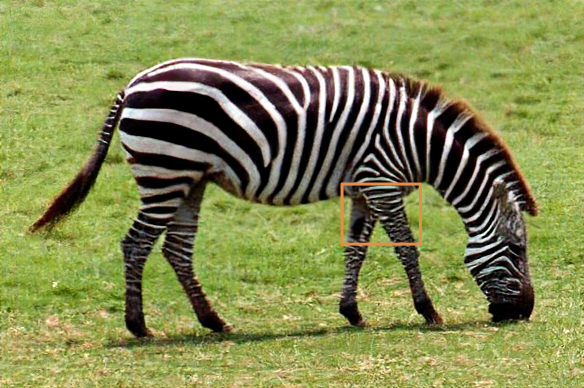}
 \includegraphics[width=2.6cm,height=1.7cm]{\epsFigPath 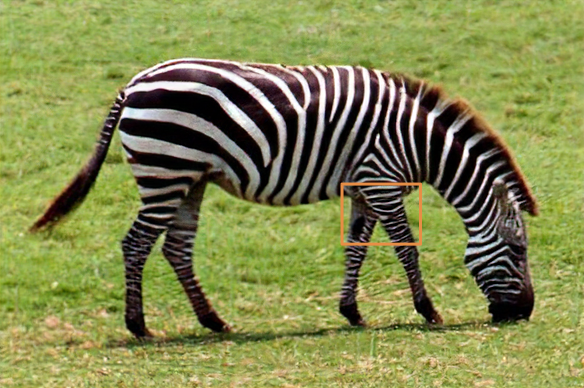}
 \includegraphics[width=2.6cm,height=1.7cm]{\epsFigPath 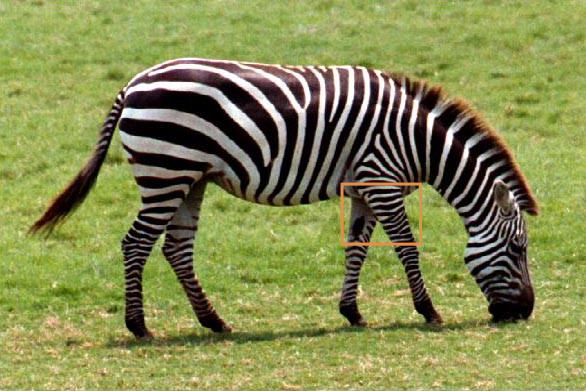}
 \includegraphics[width=2.6cm,height=1.7cm]{\epsFigPath 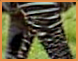}
 \includegraphics[width=2.6cm,height=1.7cm]{\epsFigPath 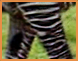}
 \includegraphics[width=2.6cm,height=1.7cm]{\epsFigPath 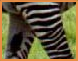}

     \caption{SISR(x4) results of the ESRGAN, the proposed LSRGAN and the ground-truth high resolution image. LSRGAN outperforms ESRGAN in structural faithfulness, details and sharpness.}
  \label{fig:open}
\end{figure}

In ESRGAN \cite{wang2018esrgan}, Wang et al. have further optimized the architecture based on the SRGAN \cite{ledig2017photo} and introduced the Residual-in-Residual Dense Block without batch normalization as the basic building unit. And the standard discriminator and generator functions are replaced by the Relativistic D and G adversarial losses $L_{D}^{Ra}$ and $L_{G}^{Ra}$ \cite{jolicoeurmartineau2018the,wang2018recovering}, which measure relative realness instead of the absolute value. The perceptual loss $L_{percep}$, is changed to be based on the pre-trained VGG19-conv54 \cite{simonyan2015very} latent space features before ReLU activation. The final G loss function is as:
\begin{equation}\label{eq:1}
  Loss_{G}^{ESR} = L_{percep}+\lambda*L_{G}^{Ra}+\eta*L_{1}
\end{equation}
The ESRGAN achieves significantly better visual quality than SRGAN and won the first place in the PIRM2018-SR Challenge. Despite this superior performance, there are still quite some artifacts in the ESRGAN results. In particular, some important structures in the restored HR images are distorted or missing when compared to the ground-truth (GT) HR images, as shown in Fig.\ref{fig:open}.

In this study, we probe for new regularization and optimization for supervised GANs to make the generator network a better approximation to the map that generates the ideal image manifold. We propose the Latent Space Regularization (LSR) and LSR-based GANs (LSRGANs) and verify them by applying them to the SISR problem. Our contributions are:
\begin{enumerate}
    \item We apply the Lipschitz continuity condition to regularize the GAN in an attempt to push the generator function to map an input sample into a manifold neighborhood that is closer to the target image. The Lipschitz condition is explicitly imposed to regularize the generator by adding a theoretically derived companion encoding network that maps images to a new optimal latent space. The encoding network is simultaneously trained with the GAN. And the Lipschitz condition is explicitly converted to new regularization terms for the generator loss function via the Karush-Kuhn-Tucker condition, and it is shown to be critical for the aforementioned encoder coupled GAN to generate good results.
    \item We verify the effect of the proposed LSR by applying it to the state of the art SISR method and the result significantly outperforms the state of the art.
    \item We propose a new SISR method by incurring a different perceptual loss based on vectorized features. It outperforms the state of the art SISR method. We further verify the applicability of the LSR by combining the LSR with the new SISR method. We find the LSR can both leverage the merits of the new SISR method and improve in many areas where the new SISR method alone incurs distortions.
\end{enumerate}
To the best of our knowledge, this is the first time the Lipschitz continuity condition is explicitly utilized to regularize and optimize the generative adversarial networks in an attempt to approximate the ideal image manifold more closely and improved results are achieved.

\section{Latent Space Regularization and GAN Optimization}
  In a CNN with ReLU as the activation function, an input can be mapped to the output by a continuous piecewise linear (PWL) function. We prove that a continuous function with a compact support can be approximated arbitrarily well by a continuous PWL function
  (see the appendix A for the proof). A CNN with enough capacity may provide a good approximation to the continuous function.
  The lower-dimensional image manifold that is embedded in the ambient image space may be learnt and represented approximately by the parameterized manifold defined by the CNN.

In supervised GAN training, as when used in the SISR problem \cite{ledig2017photo}, a noisy latent sample $z$ corresponds to a target ambient space sample $y$. The goal is to make every generated point $G(z)$ to be close to the corresponding $y$ as much as possible. We try to explore the optimal generator $G$ that can best map a sample $z$ in the latent space to a generated sample $G(z)$ in the image space so that $G(z)$ is located in a neighborhood that is close to the target image $y$ as much as possible, i.e.
\begin{equation}\label{eq:2}
  |G(z)-y|_{1} < \epsilon.
\end{equation}
We want $\epsilon$ to be small and become smaller and smaller as the training goes on.

In our design, $G$ is a CNN with the ReLU activation. It is a continuous PWL function with a compact support; and we prove it is globally Lipschitz continuous (see the appendix B for the proof).
That is, there exists a constant $K > 0$, for any latent space variables $z_1$ and $z_2$,
\begin{equation}
|G(z_1) - G(z_2)| \leq K*|z_1-z_2|.
\end{equation}

We propose to incur an encoder $L$ so that the Lipschitz Continuity Condition (LCC) can be applied in the encoded and more regularized latent space as shown in Equation (\ref{eq:3}). Note that directly bounding the difference between $G(z_1)$ and $G(z_2)$ by the difference in the original $z$ space is not a good idea because $z$ is usually corrupted by random noise or other impairments in real world applications, and the difference in the $z$ space is hard to minimize. We intend to utilize the encoder $L$ to map the ambient space to a more regularized latent space, which can then be optimized to enforce local invariance of $G$.
\begin{equation}\label{eq:3}
  |G(z)-y|_{1} \leq K*|L(G(z))-L(y)|_{1}
\end{equation}
This is a good approximation under the assumption that the set of natural HR images $\{y_i\}_i$ are in a manifold and there is a generator $G$ that can represent the manifold well enough, i.e for every $y$, there exists a good approximation $G(z_0)$. Equation (\ref{eq:3}) is then an approximation of the following equation.
\begin{equation}\label{eq:4}
  |G(z)-G(z_0)|_{1} \leq K*|L(G(z))-L(G(z_0))|_{1}
\end{equation}

We can make $G$ converge to a better approximation to the ideal mapping if we
require the left hand side (LHS) of (\ref{eq:3}) be upper bounded by a constant multiple of the regularized latent space difference (i.e. by the right hand side (RHS)) and make the RHS smaller.

Recall that the standard GAN tries to solve the following min-max problem:
\begin{equation}\label{eq:10}
\begin{split}
  &(D^{*}, G^{*}) = \\
  & \min_{G}\max_{D} (E_{y} (log D(y)) + E_{z}(1-D(G(z))))
\end{split}
\end{equation}
where $E_{y}$ and $E_{z}$ are the expectations w.r.t. the real data and the input sample distributions.
With the LCC constraint in equation (\ref{eq:3}), we can formulate the generator optimization problem as
\begin{equation}\label{eq:11}
\begin{split}
     G^{*} & =\min_{G} E_{z}(1-D(G(z))), \\
    s.t. &  \   \   E_z |y-G(z)|_{1} \leq K*E_z|L(y)-L(G(z))|_{1}
\end{split}
\end{equation}
From the Karush-Kuhn-Tucker (KKT) condition \cite{Karush2014Minima,kuhn1951nonlinear,Stephen2006Convex}, a necessary condition for the solution of the problem (\ref{eq:11}) is that it is the solution of the following optimization problem:
\begin{equation}\label{eq:12}
\begin{split}
  & G^{*} = \min_{G} E_{z}\{(1-D(G(z)))\\
  & +\eta* ( |y-G(z)|_{1} - K*|L(y)-L(G(z))|_{1})\},
\end{split}
\end{equation}
where $\eta \geq 0$ is the KKT multiplier.

Without the knowledge of the Lipschitz constant K, we make it an independent hyper parameter and further convert the problem in  (\ref{eq:12}) to
\begin{equation}\label{eq:13}
\begin{split}
   G^{*} = & \min_{G} E_{z}\{(1-D(G(z)))+\eta* |y-G(z)|_{1}\\
  & - \mu*|L(y)-L(G(z))|_{1}\}
\end{split}
\end{equation}
where $\eta \geq 0$ and $\mu \geq 0$ are independent hyper-parameters.

The above deduction can be similarly done by replacing the adversarial items with the new ones when a non standard adversarial metric such as the Relativistic discriminator \cite{jolicoeurmartineau2018the,wang2018recovering} is used.

Mathematically, equation (\ref{eq:13}) is a necessary condition to enforce the LCC in equation (\ref{eq:3}). A new GAN architecture is proposed accordingly in the following section to reduce the LHS of equation (\ref{eq:3})

\section{The LSRGAN Models and Architectures}
\begin{figure}
  \centering
  \includegraphics[width=8.3cm,height=2.5cm]{\epsFigPath 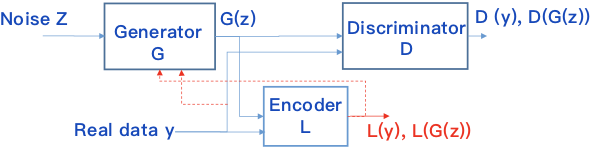}
  \caption{The proposed LSRGAN: a companion encoder $L$ is added to provide new regularization to the GAN. The red dashed line from $L$ to $G$ indicates the output from the $L$ is used to regularize the generator $G$ as part of the $G$ loss function; it is not fed through $G$ to generate new samples. So is with the red dashed line from $y$ to $G$. }
  \label{fig:diagram}
\end{figure}

The equation (\ref{eq:13}) naturally leads to a few key points for our GAN design:

	First, a companion encoding network L that maps the ambient space manifold to the latent space is augmented to the GAN. This is shown Fig. \ref{fig:diagram}. $L$ receives signals from the output of the generator $G(z)$ as well as the target data $y$. It is optimized simultaneously with the $D$ and $G$, and its outputs $L(G(z))$ and $L(y)$ are utilized to regularize the generator G. The loss function of $L$ can simply be the $L_{1}$-norm as:
\begin{equation}\label{eq:14}
  Loss_{L} = E_{z}|L(y)-L(G(z))|_{1}
\end{equation}
	Second, the generator $G$ is now also regularized by the latent space term $E_{z} |L(y)-L(G(z))|_{1}$ with an independent multiplier $\mu$. We denote it as the Latent Space Regulation (LSR). It plays a critical role to force the generator to produce sharp details that are more faithful to the targets.

Third, the resemblance of the generated sample $G(z)$ and the target $y$ is now naturally reflected by the term $E_{z} |y-G(z)|_{1}$. It is shown to be an indispensable regularization in our derivation although it is intuitive to have. In many existing GAN based SISR solutions, this term is usually deemed as a cause of soft and blurry results. We will show that sharp details are generated when it is combined with the LSR term, as in the equation (\ref{eq:13}).

We denote this GAN model the LSRGAN. And it forms the base for the following investigations to verify that the LSR helps to push the generator function to map an input sample into a manifold neighborhood that is closer to the target image.
The SISR is a good problem to which to apply the LSRGAN. Since the ESRGAN \cite{wang2018esrgan} gives the state of the art results, we would like to verify our concept and architecture on top of the ESRGAN by adding the $L$, imposing the LSR to the $G$ while keeping $D$ the same. This can be expressed as:
\begin{equation}\label{eq:15}
\begin{split}
  & Loss_{G}^{LSR} = E_{z}\{L_{percep}+\lambda*L_{G}^{Ra}\\
  & +\eta*|y-G(z)|_{1}-\mu*|L(y)-L(G(z))|_{1}\}
\end{split}
\end{equation}
And we would also like to see if the LSR works well with different perceptual loss measures other than the original $L_{percep}$. For this we introduce the cosine similarity, which measures the directional similarity between two vectors. Similar to Mechrez and et al. \cite{mechrez2018the,mechrez2018learning}, the contextual data $Gz$ and $Y$ consist of the $N$ points in the VGG19-conv34 feature maps \cite{simonyan2015very} for the images $G(z)$ and $y$. We then define the Cosine Contextual loss between $G(z)$ and $y$ as
\begin{equation}\label{eq:16}
\begin{split}
  & CCX(G(z), y) = -log(\frac{1}{N} \sum_{j}max_{i}A_{ij}),\\
  &  A_{ij} = \frac{e^{(1-d_{ij}^{'}/h)}}{\sum_{k}e^{(1-d_{ik}^{'}/h)}}, \\
  &  d_{ij}^{'} = \frac{d_{ij}}{min_{k}d_{ik}+\epsilon},\\
  &  d_{ij} = \frac{(x_{i}-r)\cdot(y_{j}-r)}{\parallel x_{i}-r\parallel_{2}\times\parallel y_{j}-r\parallel_{2}},
\end{split}
\end{equation}
where $h > 0$ is a bandwidth parameter, $\epsilon=10^{-5}$, and $r$ is a reference, e.g. the average of the points in $Y$.

We replace  the $L_{percep}$ in (\ref{eq:1}) and (\ref{eq:15}) with $CCX$ and get two new models: one with the generator function in equation (\ref{eq:20}) (denoted as CESRGAN), and the other in equation (\ref{eq:21}) (denoted as CLSRGAN).
\begin{equation}\label{eq:20}
\begin{split}
  & Loss_{G}^{CESR} = E_{z}\{CXX(G(z), y)+\lambda*L_{G}^{Ra}\\
  & +\eta*|y-G(z)|_{1}\}
\end{split}
\end{equation}
\begin{equation}\label{eq:21}
\begin{split}
  & Loss_{G}^{CLSR} = E_{z}\{CXX(G(z), y)+\lambda*L_{G}^{Ra}\\
  & +\eta*|y-G(z)|_{1}-\mu*|L(y)-L(G(z))|_{1}\}
\end{split}
\end{equation}

\subsection{Network architecture}
The CESRGAN adopts the same architecture as the ESRGAN \cite{wang2018esrgan}. They are trained using the same training algorithm. The LSRGAN and CLSRGAN share the same architecture, with the same encoder network architecture for the newly added $L$. Their $G$ and $D$ model architectures are the same too, as in the ESRGAN model. The training algorithm is similar to that of the ESRGAN, except that the companion encoder $L$ needs to be trained simultaneously.
In our implementation, the encoder $L$ is adapted from the first few layers of the VGG16 \cite{simonyan2015very} by removing the batch normalization and is followed by a few upscaling layers so that its output matches the size of the LR image that is fed to the $G$. This makes the encoder $L$ output in the same latent space as the noisy sample z. And the $L$ regularizes the latent space by minimizing the distance defined in equation (\ref{eq:14}). The $L$ is not required to be an autoencoder that would attempt to output samples that look real. In our following experiments, the $L$ is first pre-trained separately to be close to some target LR images. This is just to speed up the formal GAN training or fine-tuning, in which the $L$ parameters are only further fine-tuned to minimize the $Loss_L$ and its output is no longer required to match any target LR image.
More flexibility is also allowed to choose the $L$ architecture. We speculate the encoder network that embeds the HR image space to the LR image space may support only part of the natural image topologies, and an encoder DNN that better represents the ambient space image manifold in the latent space may produce good results for a wider range of natural images.

\section{Experiments}
\subsection{Training Details and Data}
All experiments are performed with an upscaling factor of 4 in both the horizontal and vertical directions between the LR and HR images. The DIV2K dataset \cite{agustsson2017ntire} is used for training. It contains 800 high-quality 2K-resolution images. They are flipped and rotated to augment the training dataset. HR patches of size 128x128 are cropped. The RGB channels are used as the input. The LR images are obtained by down-scaling from the HR images using the MATLAB bicubic kernel. The mini-batch size is set to 16. We implement our models in PyTorch running on NVIDIA 2080Ti GPUs.

The training process includes two stages. First, we pre-train the GAN and $L$ as PSNR-oriented models to get the initial weights of the networks. The $G$ maps the LR images to the HR images, and the $L$ maps the HR images to LR images with the $L_{1}$ loss. The Adam optimizer is used by setting $\beta_{1}=0.9$, $\beta_{2}=0.999$ and $\epsilon=10^{-8}$, without weight decaying. The learning rate is initialized as $2 \times 10^{-4}$ and decayed by a half every $2 \times 10^5$ of mini-batch updates. We train the models over 500000 iterations, until they converge. We then jointly fine-tune the $D$, $G$ and/or $L$ for the CESRGAN, LSRGAN and CLSRGAN models, with $\lambda = 5 \times 10^{-3}$, $\eta = 10^{-2}$, and $\mu = 10^{-3}$. The learning rate is set to $1 \times 10^{-4}$ and halved after [50k, 100k, 200k, 300k] iterations. We again use the Adam optimizer with the same $\beta_1$, $\beta_2$ and $\epsilon$. We alternately update the $D$, $G$, and $L$ until the models converge, or up to 500000 iterations.

\begin{table*}[!htb]
  \caption{The average PI, SSIM and PSNR(dB) values for the four test data sets for the ESR and LSR GANs. Note PI is better with a lower value. The last two columns of are the PSNR standard deviations (StdDev).}
  \label{Comparisons15}
  \centering
  \begin{tabular}{lllllllllllll}
    \toprule
    & \multicolumn{3}{c}{PI} & \multicolumn{3}{c}{SSIM} & \multicolumn{3}{c}{PSNR} & \multicolumn{2}{c}{StdDev} \\
    \cmidrule(r){2-4}\cmidrule(r){5-7}\cmidrule(r){8-10}\cmidrule(r){11-12}
          & ESR & LSR & \%change & ESR & LSR & \%change & ESR & LSR & change & ESR & LSR \\
    \midrule
    Set14 & 2.926 & \textbf{2.907} & -0.65\% & 0.718 & \textbf{0.724} & 0.84\% & 26.28 & \textbf{26.46} & 0.18dB & 4.132 & 3.913\\
    PIRM & 2.436 & \textbf{2.096} & -13.96\% & 0.669 & \textbf{0.688} & 2.84\% & 25.04 & \textbf{25.47} & 0.43dB & 3.325 & 3.144\\
    Urban100 & 3.771 & \textbf{3.520} & -6,66\% & 0.749 & \textbf{0.757} & 1.07\% & 24.36 & \textbf{24.73} & 0.37dB & 4.310 & 4.095 \\
    BSD & 2.479 & \textbf{2.388} & -3.67\% & 0.673 & \textbf{0.680} & 1.04\% & 25.32 & \textbf{25.52} & 0.20dB & 3.855 & 3.795 \\

    \bottomrule
  \end{tabular}
\end{table*}

We experimented various values for the hyper-parameter $\mu$. We find a value of $\mu$ in the range of $[0, 10^{-2}$] generally helps get better generated images. For example, $10^{-7}$ gives very sharp details that can sometimes be excessive and $10^{-3}$ gives more balanced results for LSR GANs
(see the supplementary for experimental results).
$10^{-3}$ is used for training the LSRGAN and CLSRGAN in the following experiments.

\subsection{Evaluation Results}
We evaluate the models on widely used benchmark datasets: 
Set14 \cite{zeyde2010on}, BSD100 \cite{martin2001a}, Urban100 \cite{huang2015single} and the PIRM test dataset that is provided in the PIRM-SR Challenge \cite{DBLP:journals/corr/abs-1809-07517}.

We performed experiments to check the effects of imposing the LSR to the ESRGAN and CESRGAN models. The purpose is to verify that the proposed LSR and encoder $L$-coupled GAN architecture can push the generator to produce a sample that is closer to the ground truth than those of the GANs without the LSR and encoder $L$.

We measure the values of the PSNR, SSIM and Perceptual Index (PI) \cite{DBLP:journals/corr/abs-1809-07517} for each model. It is recognized in the research community that numerical scores such as PSNR and/or SSIM are not suitable for differentiating and evaluating image perceptual quality because they do not correlate very well with image subjective quality \cite{DBLP:journals/corr/abs-1809-07517,ledig2017photo}, PI was devised to overcome part of this deficiency\cite{DBLP:journals/corr/abs-1809-07517}. We adopt it as the main check point along with subjective quality check. Subjective quality check is to compensate the lack of effective numeric measures for perceptual quality. We present some representative qualitative results. We will check how well the internal image structures of the generated samples match the ground truth images and how details and sharpness look.

\begin{figure}[!htb]

 \includegraphics[width=2.7cm,height=2.0cm]{\epsFigPath 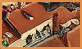}
 \includegraphics[width=2.7cm,height=2.0cm]{\epsFigPath 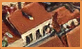}
 \includegraphics[width=2.7cm,height=2.0cm]{\epsFigPath 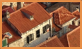}

 \includegraphics[width=2.7cm,height=2.0cm]{\epsFigPath 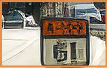}
 \includegraphics[width=2.7cm,height=2.0cm]{\epsFigPath 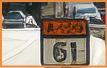}
 \includegraphics[width=2.7cm,height=2.0cm]{\epsFigPath 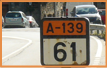}

 \includegraphics[width=2.7cm,height=2.0cm]{\epsFigPath 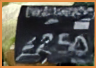}
 \includegraphics[width=2.7cm,height=2.0cm]{\epsFigPath 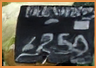}
 \includegraphics[width=2.7cm,height=2.0cm]{\epsFigPath 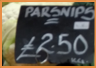}

 \includegraphics[width=2.7cm,height=2.2cm]{\epsFigPath 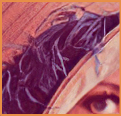}
 \includegraphics[width=2.7cm,height=2.2cm]{\epsFigPath 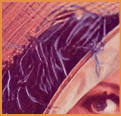}
 \includegraphics[width=2.7cm,height=2.2cm]{\epsFigPath 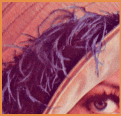}

 \includegraphics[width=2.7cm,height=2.0cm]{\epsFigPath 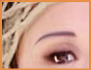}
 \includegraphics[width=2.7cm,height=2.0cm]{\epsFigPath 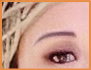}
 \includegraphics[width=2.7cm,height=2.0cm]{\epsFigPath 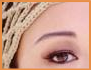}

 \includegraphics[width=2.7cm,height=2.0cm]{\epsFigPath zebra_ESR_W3.png}
 \includegraphics[width=2.7cm,height=2.0cm]{\epsFigPath zebra_LSR_w3.png}
 \includegraphics[width=2.7cm,height=2.0cm]{\epsFigPath zebra_Orig_W3.png}

 \includegraphics[width=2.7cm,height=2.0cm]{\epsFigPath 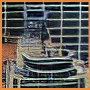}
 \includegraphics[width=2.7cm,height=2.0cm]{\epsFigPath 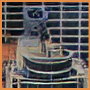}
 \includegraphics[width=2.7cm,height=2.0cm]{\epsFigPath 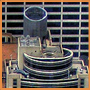}

  \caption{Results from LR to HR (4x) experiments.
 From left to right are patches from the generated images of ESR, LSR GANs and the HR ground truth.
 }
  \label{fig:patches15}
\end{figure}

\begin{figure*}[!htb]
  \centering
  \includegraphics[width=3.4cm,height=2.2cm]{\epsFigPath 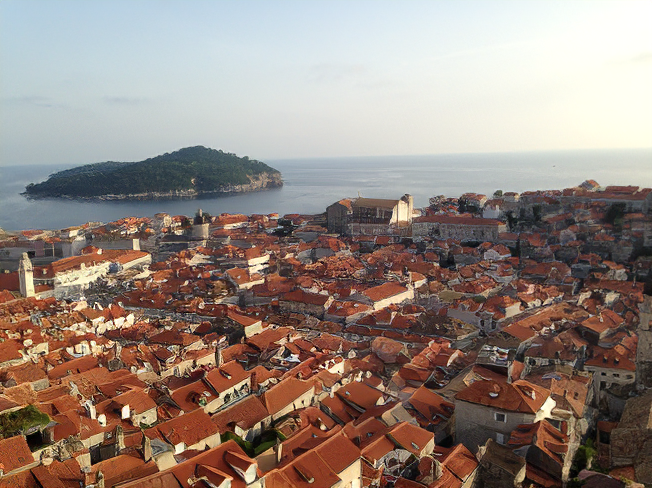}
  \includegraphics[width=3.4cm,height=2.2cm]{\epsFigPath 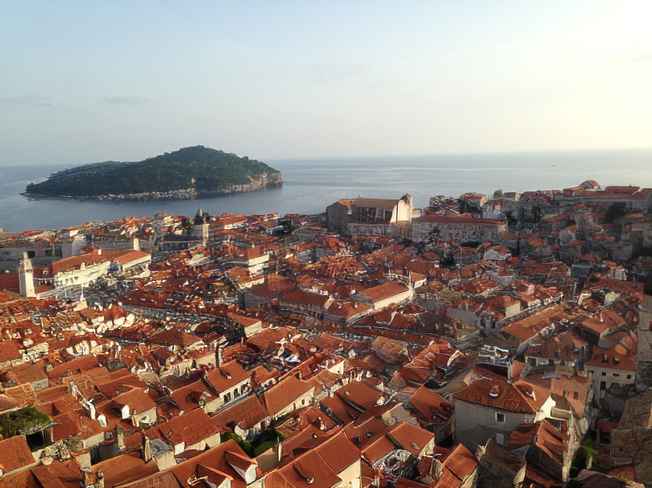}
  \includegraphics[width=3.4cm,height=2.2cm]{\epsFigPath 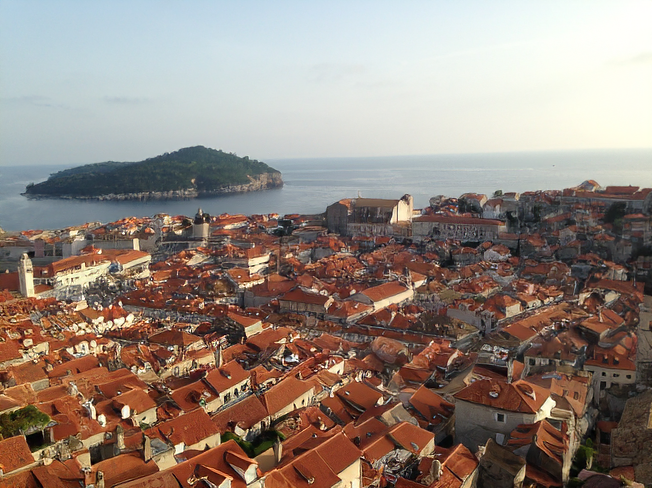}
  \includegraphics[width=3.4cm,height=2.2cm]{\epsFigPath 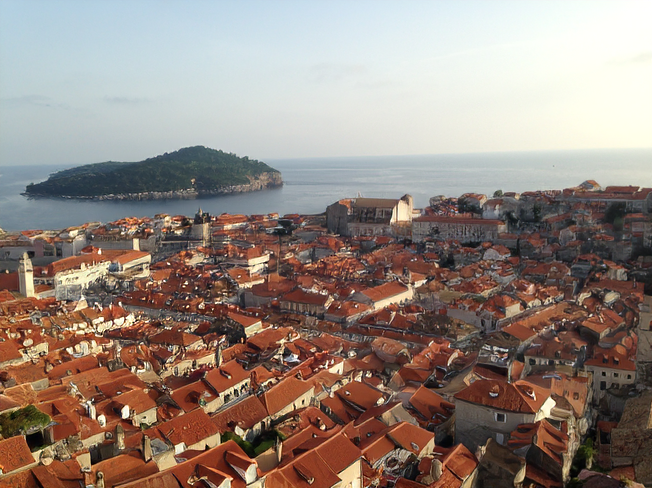}
  \includegraphics[width=3.4cm,height=2.2cm]{\epsFigPath 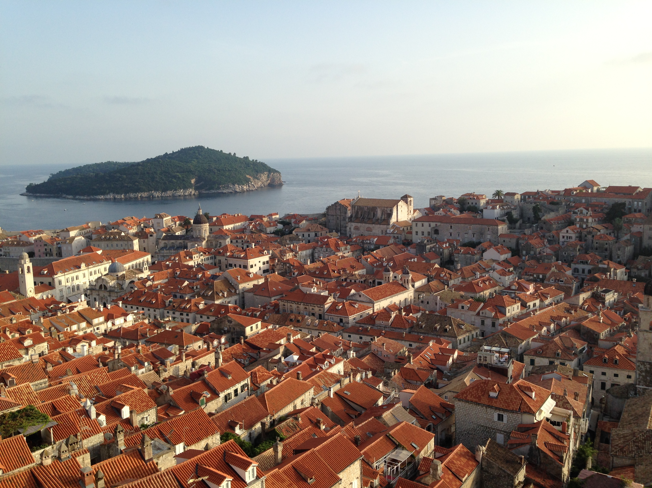}

  \includegraphics[width=3.4cm,height=2.2cm]{\epsFigPath 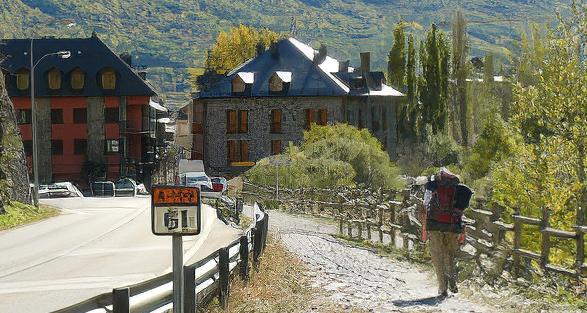}
  \includegraphics[width=3.4cm,height=2.2cm]{\epsFigPath 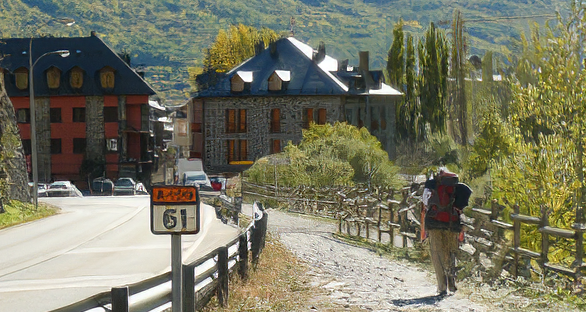}
  \includegraphics[width=3.4cm,height=2.2cm]{\epsFigPath 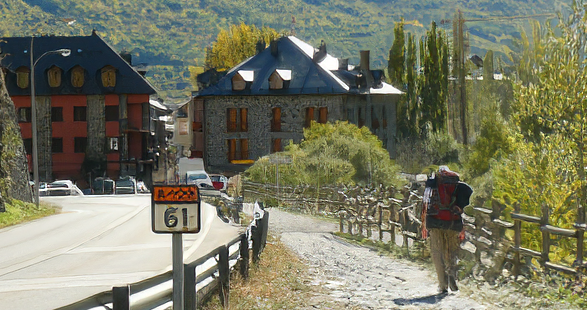}
  \includegraphics[width=3.4cm,height=2.2cm]{\epsFigPath 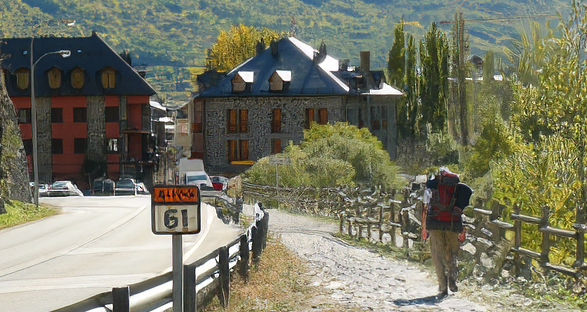}
  \includegraphics[width=3.4cm,height=2.2cm]{\epsFigPath 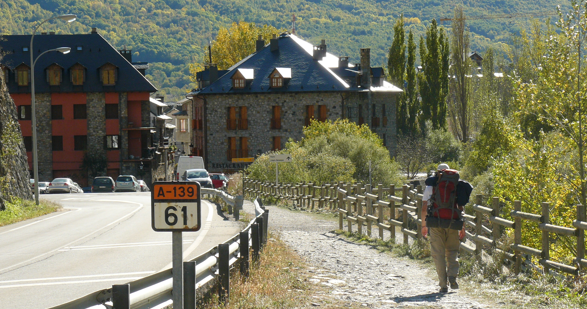}

  \includegraphics[width=3.4cm,height=2.2cm]{\epsFigPath 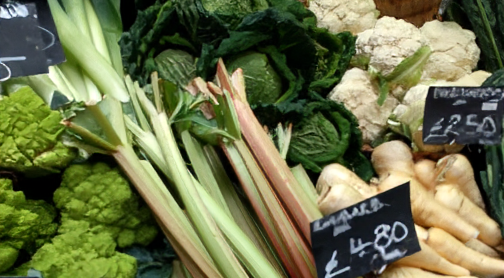}
  \includegraphics[width=3.4cm,height=2.2cm]{\epsFigPath 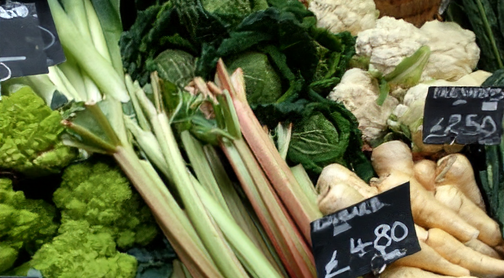}
  \includegraphics[width=3.4cm,height=2.2cm]{\epsFigPath 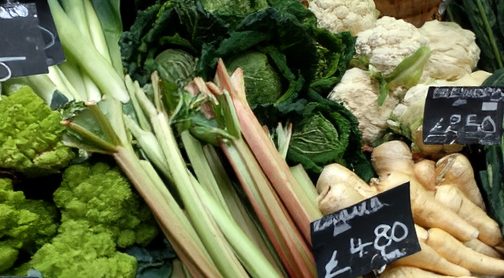}
  \includegraphics[width=3.4cm,height=2.2cm]{\epsFigPath 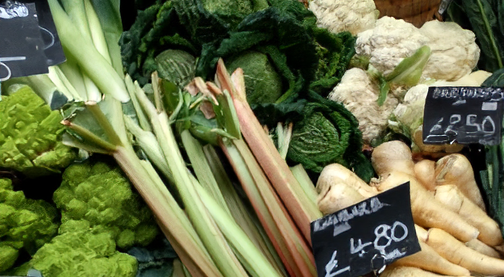}
  \includegraphics[width=3.4cm,height=2.2cm]{\epsFigPath 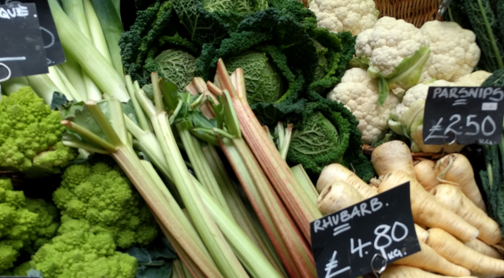}

  \includegraphics[width=3.4cm,height=3.4cm]{\epsFigPath 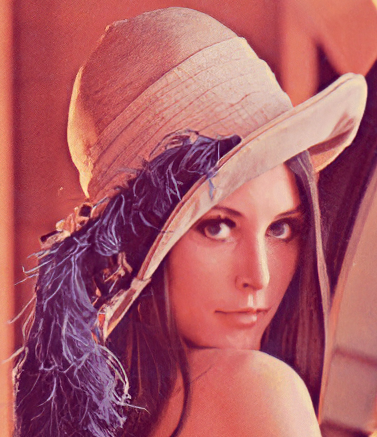}
  \includegraphics[width=3.4cm,height=3.4cm]{\epsFigPath 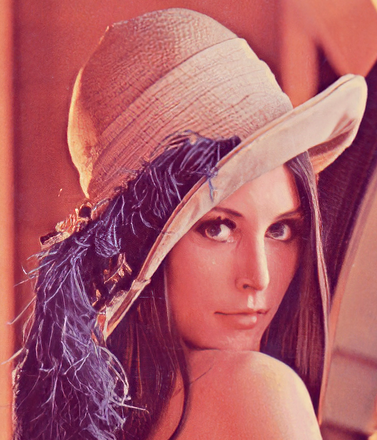}
  \includegraphics[width=3.4cm,height=3.4cm]{\epsFigPath 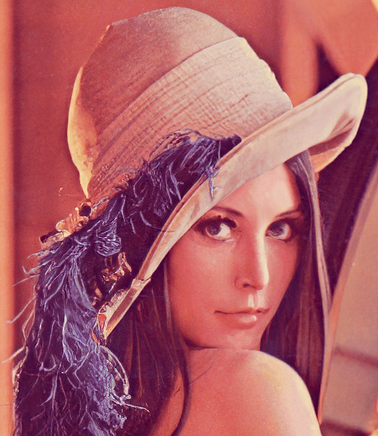}
  \includegraphics[width=3.4cm,height=3.4cm]{\epsFigPath 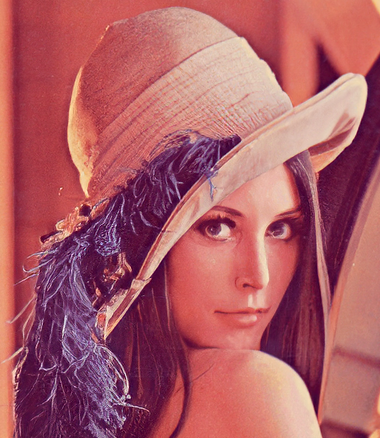}
  \includegraphics[width=3.4cm,height=3.4cm]{\epsFigPath 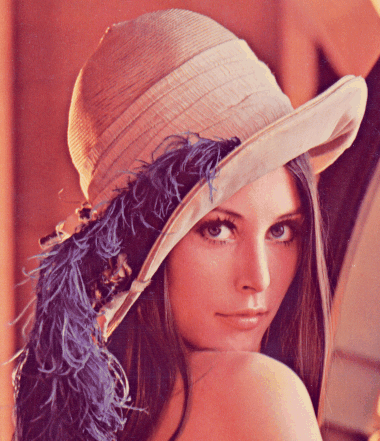}

  \includegraphics[width=3.4cm,height=3.5cm]{\epsFigPath 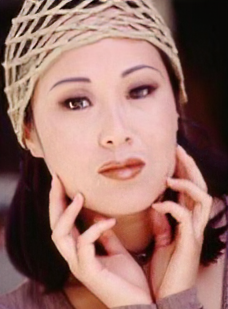}
  \includegraphics[width=3.4cm,height=3.5cm]{\epsFigPath 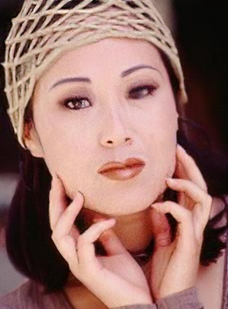}
  \includegraphics[width=3.4cm,height=3.5cm]{\epsFigPath 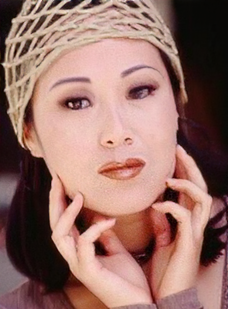}
  \includegraphics[width=3.4cm,height=3.5cm]{\epsFigPath 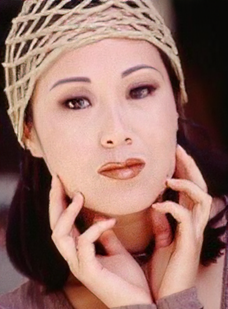}
  \includegraphics[width=3.4cm,height=3.5cm]{\epsFigPath 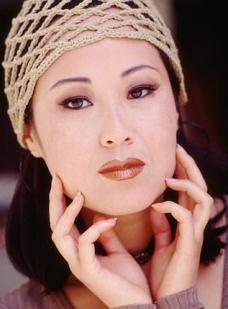}

  \includegraphics[width=3.4cm,height=2.2cm]{\epsFigPath 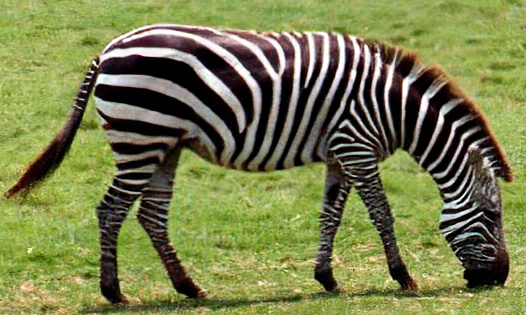}
  \includegraphics[width=3.4cm,height=2.2cm]{\epsFigPath 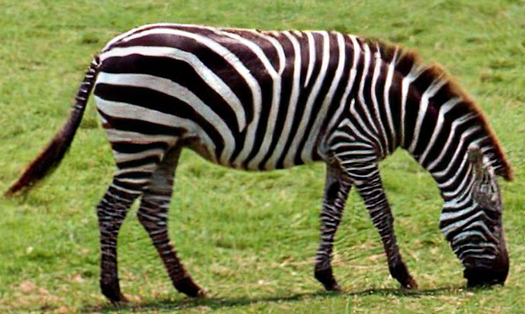}
  \includegraphics[width=3.4cm,height=2.2cm]{\epsFigPath 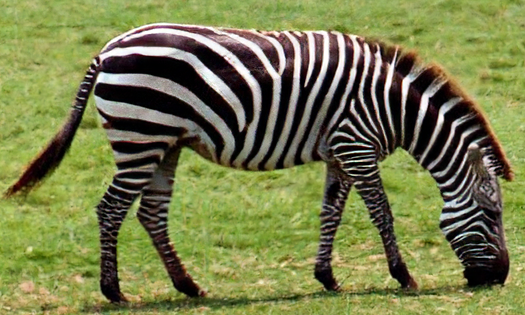}
  \includegraphics[width=3.4cm,height=2.2cm]{\epsFigPath 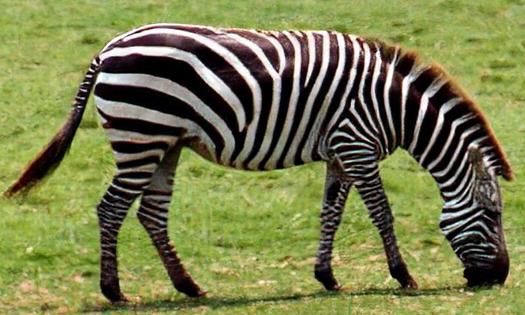}
  \includegraphics[width=3.4cm,height=2.2cm]{\epsFigPath 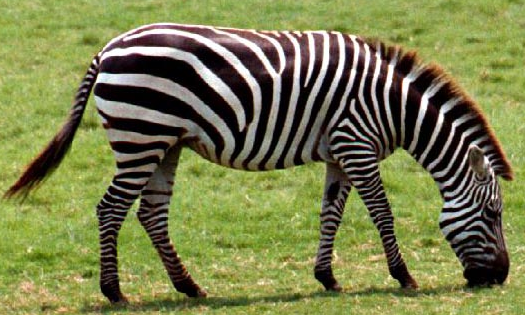}

     \caption{Results from LR to HR (4x) experiments, full images. Shown from the left to the right are the results from the ESR, LSR, CESR, CLSR models and the GT HR images. LSR and CLSR GANs outperform ESR amd CESR GANs in structure faithfulness, sharpness and detail clarity.}
  \label{fig:wholes}
\end{figure*}

Comparison data between the LSR and ESR are listed in Table \ref{Comparisons15}. Some representative local patch images are shown in Fig. \ref{fig:patches15}. A few of the full images are shown in Fig. \ref{fig:wholes} (a few are cropped around the center areas for better viewing). Full images are best for viewing when being zoomed in.

\begin{table*}[!htb]
  \caption{The average PI, SSIM and PSNR(dB) values for the four test data sets for the CESR and CLSR GANs. Note PI is better with a lower value. The last two columns of are the PSNR standard deviations (StdDev).}
  \label{Comparisons16}
  \centering
  \begin{tabular}{llllllllll}
    \toprule
    & \multicolumn{2}{c}{PI} & \multicolumn{2}{c}{SSIM} & \multicolumn{2}{c}{PSNR} & \multicolumn{2}{c}{StdDev} \\
    \cmidrule(r){2-3}\cmidrule(r){4-5}\cmidrule(r){6-7}\cmidrule(r){8-9}
          & CESR & CLSR & CESR & CLSR & CESR & CLSR & CESR & CLSR\\
    \midrule
    Set14 & \textbf{2.738} & 2.820 & 0.725 & \textbf{0.731} & 26.43 & \textbf{26.52} & 3.887 & 3.720\\
    PIRM & 2.117 & \textbf{2.112} & 0.687 & \textbf{0.692} & 25.45 & \textbf{25.63} & 3.104 & 3.096\\
    Urban100 & 3.513 & \textbf{3.511} & 0.758 & \textbf{0.760} & 24.71 & \textbf{24.76} & 4.210 & 4.202 \\
    BSD & 2.311 & \textbf{2.290} & 0.680 & \textbf{0.680} & 25.52 & \textbf{25.53} & 3.770 & 3.787 \\
    \bottomrule
  \end{tabular}
\end{table*}

First, we can see that the LSR improves all the average PSNR, SSIM and PI scores over ESR, with the PI (which emphasizes perceptual quality) and PSNR being improved more significantly. Next we compare the qualitative quality of the generated images. LSR makes the internal structures more faithful to the GT HR images in most cases. For example, the house structure looks more right in the LSR image (the second image from the left in the first row of images in Fig. \ref{fig:patches15}) than in the ESR image (the first image from the left in the first row); the digits and cars in the LSR images look more solid in the second and third rows; the hat decor and eyes in the LSR images are sharper and fine details are more separated in the fourth and fifth rows; the zebra leg stripes and fine building structures in the LSR images are better disentangled in the sixth and seventh rows. The chimney nearly disappears and color does not look right in many places in the ESR result in the seventh row but they look more correct in the LSR result.

\begin{figure}[!htb]
  \centering

 \includegraphics[width=2.7cm,height=2.0cm]{\epsFigPath 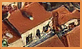}
 \includegraphics[width=2.7cm,height=2.0cm]{\epsFigPath 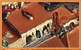}
 \includegraphics[width=2.7cm,height=2.0cm]{\epsFigPath 204_Orig_W3.png}

 \includegraphics[width=2.7cm,height=2.0cm]{\epsFigPath 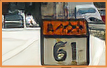}
 \includegraphics[width=2.7cm,height=2.0cm]{\epsFigPath 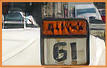}
 \includegraphics[width=2.7cm,height=2.0cm]{\epsFigPath 280_Orig_W3.png}

 \includegraphics[width=2.7cm,height=2.0cm]{\epsFigPath 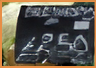}
 \includegraphics[width=2.7cm,height=2.0cm]{\epsFigPath 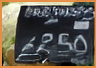}
 \includegraphics[width=2.7cm,height=2.0cm]{\epsFigPath 283_Orig_W3.png}

 \includegraphics[width=2.7cm,height=2.2cm]{\epsFigPath 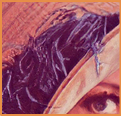}
 \includegraphics[width=2.7cm,height=2.2cm]{\epsFigPath 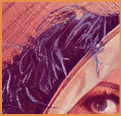}
 \includegraphics[width=2.7cm,height=2.2cm]{\epsFigPath lenna_Orig_W3.png}

 \includegraphics[width=2.7cm,height=2.0cm]{\epsFigPath 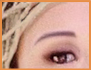}
 \includegraphics[width=2.7cm,height=2.0cm]{\epsFigPath 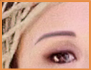}
 \includegraphics[width=2.7cm,height=2.0cm]{\epsFigPath woman_Orig_W3.png}

 \includegraphics[width=2.7cm,height=2.0cm]{\epsFigPath 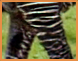}
 \includegraphics[width=2.7cm,height=2.0cm]{\epsFigPath 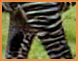}
 \includegraphics[width=2.7cm,height=2.0cm]{\epsFigPath zebra_Orig_W3.png}

 \includegraphics[width=2.7cm,height=2.0cm]{\epsFigPath 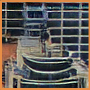}
 \includegraphics[width=2.7cm,height=2.0cm]{\epsFigPath 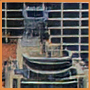}
 \includegraphics[width=2.7cm,height=2.0cm]{\epsFigPath img_073_Orig_W3.png}

  \caption{Results from LR to HR (4x) experiments.
 From left to right are patches from the generated images of CESR, CLSR GANs and the HR ground truth.
 }
  \label{fig:patches16}
\end{figure}

We also compare the CLSR with CESR to see if the LSR works well with different loss measures. The comparison data and representative local patch images are shown in Table \ref{Comparisons16} and Fig. \ref{fig:patches16} respectively. And some of the full images are shown in Fig. \ref{fig:wholes} (some of them are cropped around the center areas for better viewing).
Qualitatively, what's observed in the above comparison between LSR and ESR is also generally true for CLSR vs. CESR. See the 2nd and 1st images from the left in each row in Fig.\ref{fig:patches16}. The cars, digits, eyes and woven hat from row two to row five are obviously improved and sharper in CLSR. The average PSNR, SSIM and PI scores are still better (although marginally) for CLSR than for CESR except the PI of Set14. We notice CESR helps in many cases when compared to the ESR (see the images in the 1st column in Fig. \ref{fig:patches15}). But CESR degrades quality severely sometimes. For example, the zebra leg stripes are smeared and spurious lines penetrate the building chimney, which is no longer recognizable in the CESR result. The CLSR corrects most of these errors.

In both cases the LSR results in improved quality in general. This shows generality of the LSR to some degree. Overall, the GANs with LSR generate results that match the local image structures in the GT HR image better and sharper than the GANs without LSR.

The standard deviations for the PI, SSIM and PSNR are usually smaller in the LSR GANs except one case for the PSNR. So only the PSNR standard deviations are listed in Tables \ref{Comparisons15} and \ref{Comparisons16} (see the last columns). The numeric values resulted from the LSR GANs are generally more consistent.


\begin{table*}[!htb]
  \caption{The average $L_1$ values for the four test data sets for the CESR and CLSR GANs.}
  \label{Comparisons6}
\centering
\begin{tabular}{ |p{1cm}p{1cm}|p{1cm}p{1cm}|p{1cm}p{1cm}|p{1cm}p{1cm}|  }
 \hline
 \multicolumn{2}{|c|}{Set14} & \multicolumn{2}{|c|}{PIRM}& \multicolumn{2}{|c|}{Urban}& \multicolumn{2}{|c|}{BSD100}\\
 \hline
      CESR & CLSR & CESR & CLSR & CESR & CLSR & CESR & CLSR\\
 \hline
     0.463 & \textbf{0.444} & 0.463 & \textbf{0.429} & 0.460 & \textbf{0.436} & 0.453 & \textbf{0.420}\\
 \hline
\end{tabular}
\end{table*}

It is interesting to compare the ESR and the CESR GANs although this is not the focus of the paper.  From the images in the first columns of Figs. \ref{fig:patches15} and \ref{fig:patches16}, we can see that the CESRGAN results are better than the ESRGAN results in general. So are the numeric measures as shown in Tables \ref{Comparisons15} and \ref{Comparisons16}. The CESRGAN outperforms the state of the art ESRGAN for the SISR problem too. Careful check shows that the CESRGAN still distort the structures in many places. The obvious numeric measure improvements (in PI and PSNR, e.g.) over the ESRGAN do not translate to subjective quality improvements proportionally, unlike what the LSR GANs have achieved.

Finally, the intent of incurring the LCC constraint in equation \ref{eq:3} is to force the generator to create images that are closer to the targets in the sense that the $L_1$ distance $|G(z)-y|_{1} $ may be made smaller.
We therefore measure the average $L_1$ error on the test data sets for the CLSR and CESR GANs.
We find that average $L_1$ is lower in CLSR than in CESR in all these cases, as shown in Table \ref{Comparisons6}. The values are calculated for the Y channels of the images and are normalized to the range of $[0,1]$. This seems to indicate that LSR is effective.

The $L_1$ values for the LSR and ESR GANs are not much different and not listed. The reason might be that the ESR model (from the public domain model that is provided by the ESRGAN authors \cite{esrgan_model,wang2018esrgan}) was trained using more training data sets in addition to the DIV2K data set, which is the only data set we use for training.

\section{Discussions}
We explicitly apply the Lipschitz continuity condition to regularize the GAN by adding a coupling encoding network and by converting the Lipschitz condition to latent space regularization terms for the GAN generator via the Karush-Kuhn-Tucker condition. The GAN and the coupling latent space network are simultaneously optimized. Experiments on SISR show that optimizing the GAN via simultaneous latent space regularization and optimization pushes the generator converge to a map that generates samples that are more faithful to the desired targets and sharper. The results outperform the state of the art.

Some aspects of the model can be investigated more thoroughly.
 We have opted for a encoder network that maps the high-resolution image space to the low-resolution space.
 The encoder $L$ network that embeds the HR image manifold to the LR space may only support the part of the natural image topologies which the $L$ represents. An encoder that can better represent the natural HR or LR image distributions in the latent space may produce good results for a wider range of LR images.

The effect of the LSR varies with different adversarial terms. We find the LSR achieves better results for SISR with the Relativistic adversarial losses \cite{jolicoeurmartineau2018the} \cite{wang2018recovering} than with the standard adversarial losses. We also find that the LSR works well with the cosine similarity based contextual loss for SISR. Further investigating how different terms work with the LSR may be worthwhile.

Application-wise, the proposed LSR-based GAN models may be applied to vision tasks other than SISR, such as image restoration and image inpainting.

And for the SISR problem, we only used the 800 images of the DIV2K data set and their augmented ones for training. Including more diversified data for training may further improve the results.

We may revisit these in the future.

{\small
\bibliographystyle{aaai}
\bibliography{cites}
}

\appendix

\label{gen_inst}

\section{Proof for the Continuous Piecewise Linear Approximation Proposition}
\label{gen_inst}
In section 2 of the paper, we claim that a continuous function with a compact support can be approximated arbitrarily well by a continuous piecewise linear (PWL) function. The proof is provided here. Without loss of generality, the following analysis is performed for real functions in one dimensional space.

Let $C[a, b]$ be the set of functions that are continuous in the closed interval $[a, b]$.

 \textbf{Continuous Piecewise Linear Approximation Proposition}. Let $f$ $\in C[a, b]$. For every $\epsilon > 0$, there is a continuous piecewise linear function $f_{pwl}$ such that for every x $\in [a, b]$,
\begin{equation}\label{eq:42}
  |f(x)-f_{pwl}(x)| < \epsilon.
\end{equation}

First recall the \textbf{Weierstrass Approximation Theorem}. Let f$\in C[a, b]$. Then, for every $\epsilon > 0$, there is a polynomial p such that for every x $\in [a, b]$,
\begin{equation}
  |f(x) - p(x)| < \epsilon.
\end{equation}

Proof of the Piecewise Linear Approximation Proposition.  Let ${x_{0} = a < x_{1} < ... < x_{n} = b}$ be a sequence of $n+1$ different points in $[a, b]$. Define $f_{pwl}$ to be the continuous piecewise linear function that interpolates $f$ at the ${x_{i}}$ , i.e., for any $x \in [a, b]$, there is an index $i$ with $i = 0, 1, $ ..., or $n-1$, such that $x \in [x_{i} , x_{i+1}]$; and

\begin{equation}\label{eq:43}
  f_{pwl}(x)= f(x_{i})+ \frac{x-x_{i}}{x_{i+1}-x_{i}} (f(x_{i+1})-f(x_{i}))
\end{equation}
And we can derive that
\begin{equation}\label{eq:44}
\begin{split}
  f(x)-f_{pwl}(x) & = \frac{x_{i+1}-x}{x_{i+1}-x_{i}} (f(x)-f(x_{i}))\\
  & +\frac{x-x_{i}}{x_{i+1}-x_{i}} (f(x)-f(x_{i+1}))
\end{split}
\end{equation}
From this we can further derive that
\begin{equation}\label{eq:45}
\begin{split}
  & |f(x)-f_{pwl}(x)| \\
  & \leq \max_{x}(|f(x)-f(x_{i})|, |f(x)-f(x_{i+1})|) \\
  & \leq \sup_{(x,y)}\{|f(x)-f(y)|: x,y \in [x_{i} , x_{i+1}]\}
\end{split}
\end{equation}

By the Weierstrass Approximation Theorem, for every $\epsilon > 0$, there is a polynomial p such that for every x$\in$[a,b],
\begin{equation}
  |f(x) - p(x)| < \frac{\epsilon}{3}.
\end{equation}
Therefore, for any $x, y \in [a, b]$, we have
\begin{equation}\label{eq:46}
\begin{split} & |f(x)-f(y)| \\
    & \leq |f(x)-p(x)|+|p(x)-p(y)|+|p(y)-f(y)| \\
    & \leq \frac{2\epsilon}{3}+|p(x)-p(y)|
\end{split}
\end{equation}
The polynomial p on the compact support [a, b] is Lipschitz continuous (actually it is continuously differentiable everywhere). There is a constant $K > 0$ such that for any x, y $\in$ [a, b],
\begin{equation}
|p(x) - p(y)| \leq K*|x-y|.
\end{equation}

We sample the sequence ${x_{0} = a < x_{1} < ... < x_{n} = b}$ dense enough such that
\begin{equation}\label{eq:47}
  \max_{i=0}^{n-1} |x_{i}-x_{i+1}| \leq  \frac{\epsilon}{3K}
\end{equation}
Combining (\ref{eq:45}) to (\ref{eq:47}), the proposition (\ref{eq:42}) is proven.       \  \   \   \   \   \  \   \   \   \             

\begin{figure*}[!htb]
  \centering
  \includegraphics[width=4.2cm,height=3.1cm]{\epsFigPath 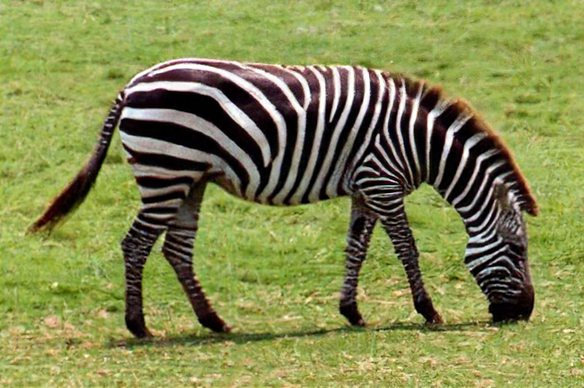}
  \includegraphics[width=4.2cm,height=3.1cm]{\epsFigPath 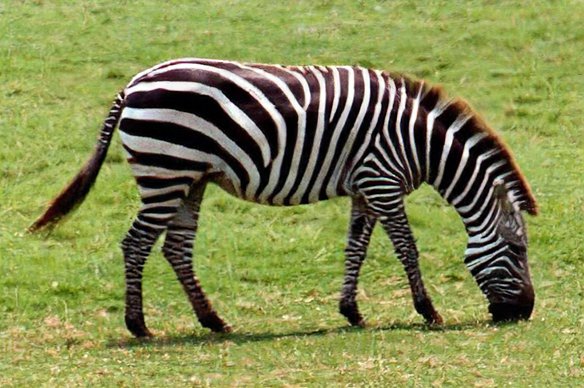}
  \includegraphics[width=4.2cm,height=3.1cm]{\epsFigPath 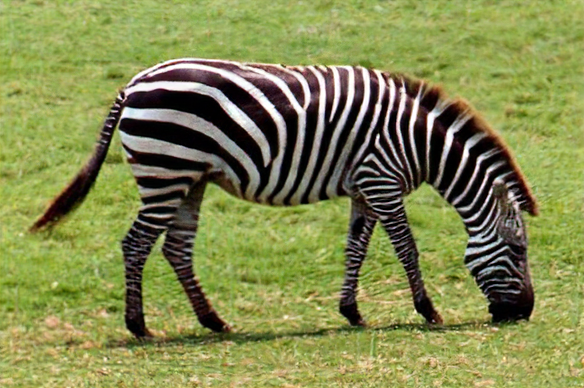}
  \includegraphics[width=4.2cm,height=3.1cm]{\epsFigPath 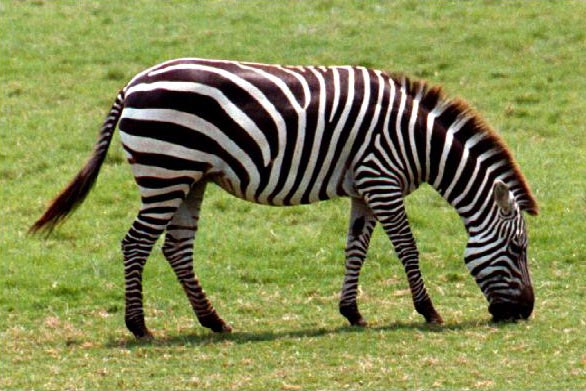}

  \includegraphics[width=4.2cm,height=3.1cm]{\epsFigPath 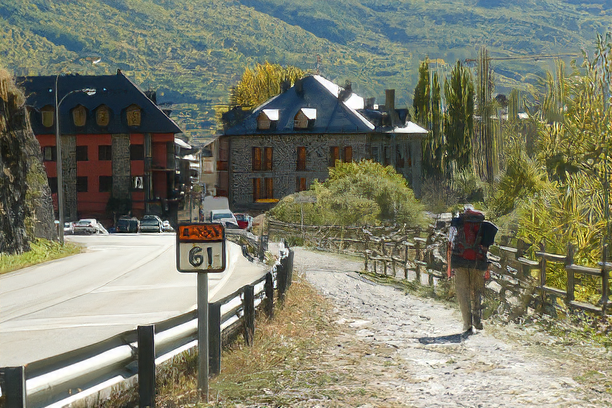}
  \includegraphics[width=4.2cm,height=3.1cm]{\epsFigPath 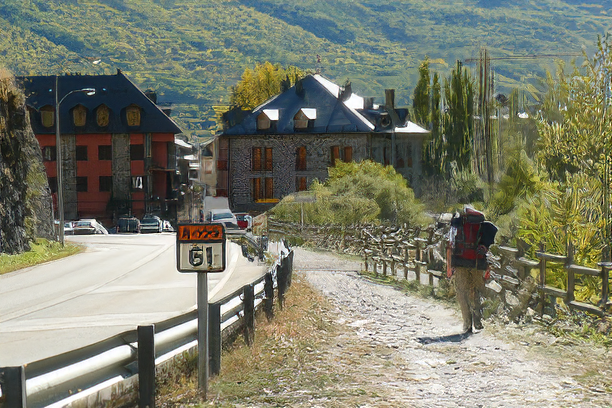}
  \includegraphics[width=4.2cm,height=3.1cm]{\epsFigPath 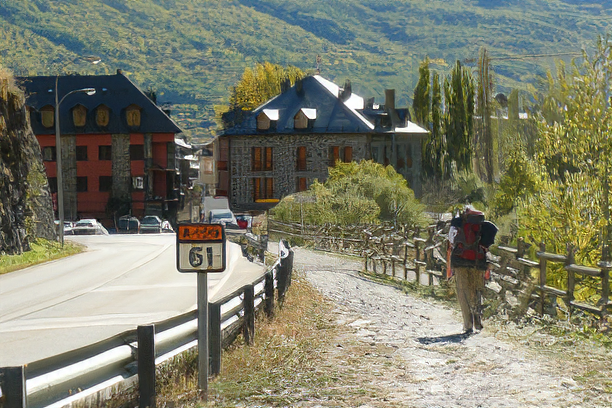}
  \includegraphics[width=4.2cm,height=3.1cm]{\epsFigPath 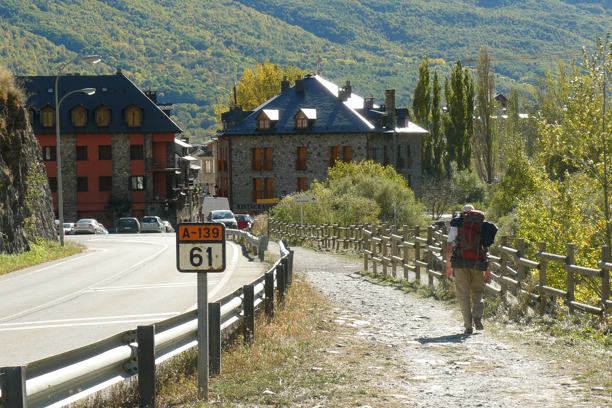}

  \includegraphics[width=4.2cm,height=4.2cm]{\epsFigPath 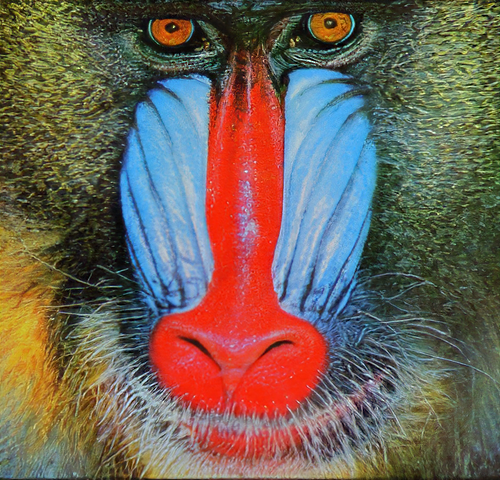}
  \includegraphics[width=4.2cm,height=4.2cm]{\epsFigPath 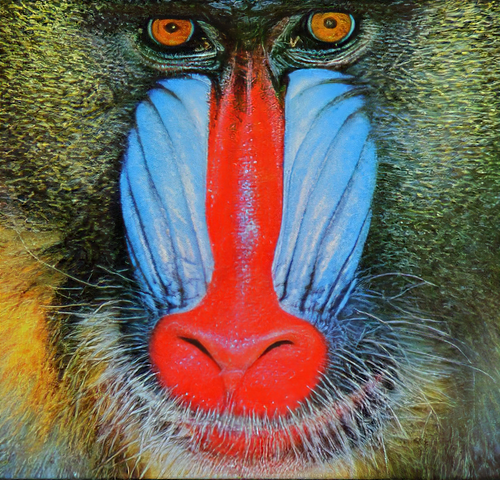}
  \includegraphics[width=4.2cm,height=4.2cm]{\epsFigPath 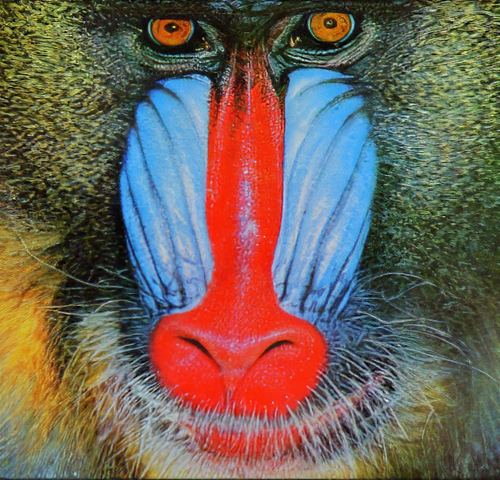}
  \includegraphics[width=4.2cm,height=4.2cm]{\epsFigPath 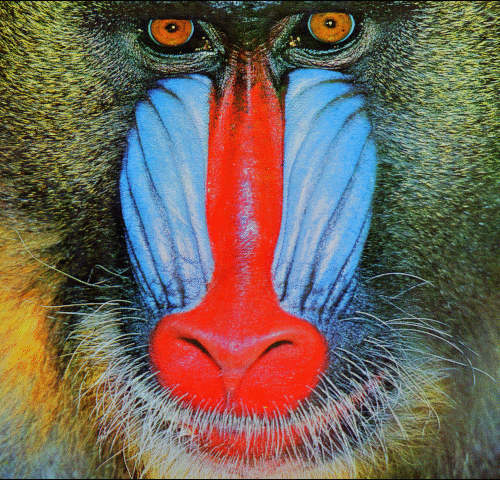}

  \includegraphics[width=4.2cm,height=4.2cm]{\epsFigPath 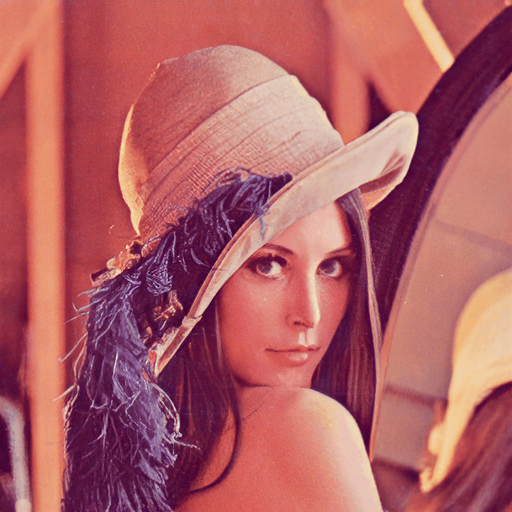}
  \includegraphics[width=4.2cm,height=4.2cm]{\epsFigPath 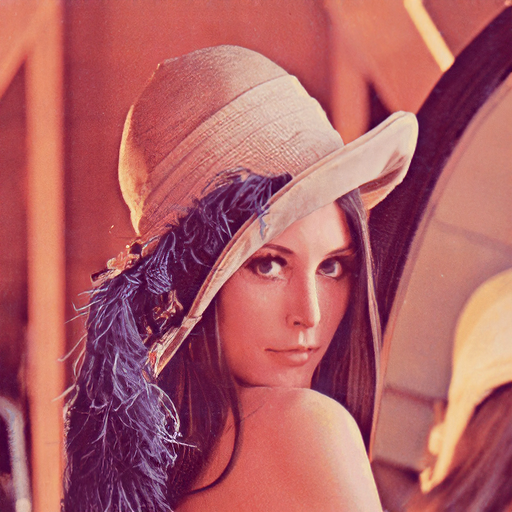}
  \includegraphics[width=4.2cm,height=4.2cm]{\epsFigPath 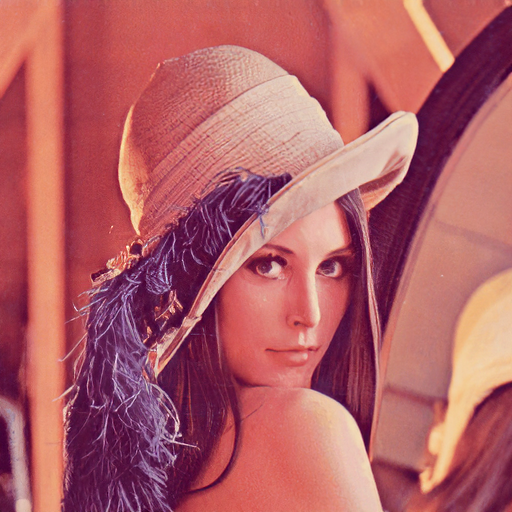}
  \includegraphics[width=4.2cm,height=4.2cm]{\epsFigPath 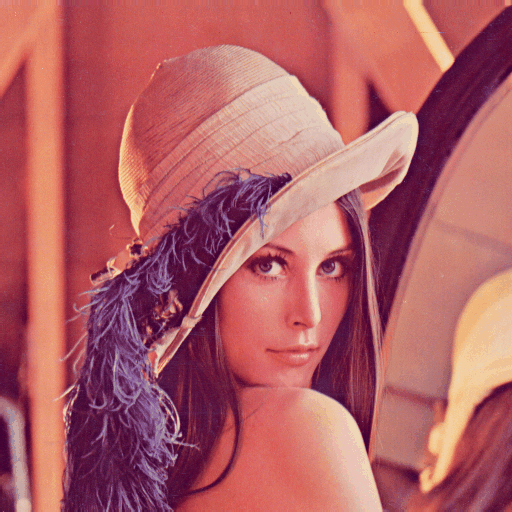}

  \includegraphics[width=4.2cm,height=3.1cm]{\epsFigPath 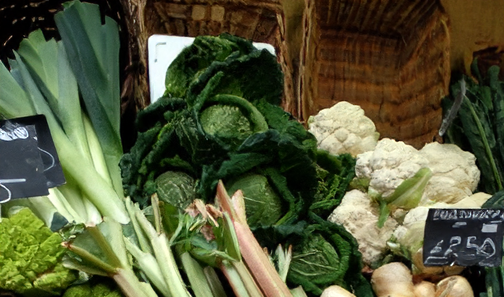}
  \includegraphics[width=4.2cm,height=3.1cm]{\epsFigPath 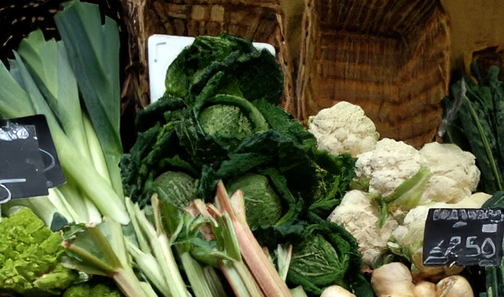}
  \includegraphics[width=4.2cm,height=3.1cm]{\epsFigPath 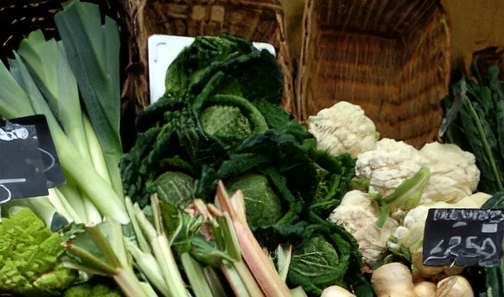}
  \includegraphics[width=4.2cm,height=3.1cm]{\epsFigPath 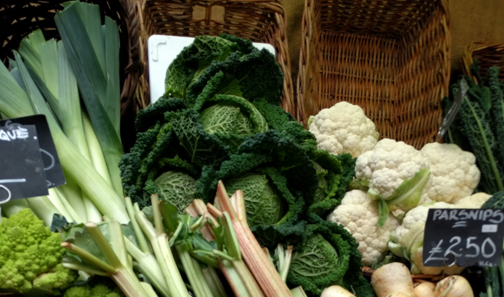}

     \caption{Results from LR to HR (4x) experiments. Shown from the left to the right are the results from the LSRGAN model with $\mu$ = 0, $10^{-7}$ and $10^{-3}$, and the ground truth HR images.}
  \label{fig:wholes2}
\end{figure*}

\begin{table*}[!htb]
  \caption{The average PSNR(dB), SSIM and PI values measured on the test datasets of Set14, PIRM SR Challenge, Urban100 and BSD100 for the LSRGAN model with the hyper parameter $\mu$ values of 0, $10^{-7}$ and $10^{-3}$. Note that PI is regarded as being better with a lower value.}
  \label{Experimental Results}
  \centering
  \begin{tabular}{lllllllllllll}
    \toprule
     & \multicolumn{3}{c}{Set14} & \multicolumn{3}{c}{PIRM-SR} & \multicolumn{3}{c}{Urban100} & \multicolumn{3}{c}{BSD100}\\
    \cmidrule(r){2-4}\cmidrule(r){5-7}\cmidrule(r){8-10}\cmidrule(r){11-13}
          & 0 & $10^{-7}$ & $10^{-3}$ & 0 & $10^{-7}$ & $10^{-3}$ & 0 & $10^{-7}$ & $10^{-3}$ & 0 & $10^{-7}$ & $10^{-3}$\\
    \midrule
    PI   & 2.908 & 2.912 & \textbf{2.907} & 2.155 & 2.159 & \textbf{2.096} & 3.551 & 3.533 & \textbf{3.520} & 2.404 & \textbf{2.374} & 2.388\\
    PSNR & 26.44 & \textbf{26.51} & 26.46 & 25.44 & 25.46 & \textbf{25.47} & 24.54 & 24.61 & \textbf{25.73} & 25.35 & 25.37 & \textbf{25.52}  \\
    SSIM & 0.723 & \textbf{0.727} & 0.724 & 0.683 & 0.685 & \textbf{0.688} & 0.751 & 0.756 & \textbf{0.757} & 0.671 & 0.673 & \textbf{0.680} \\
    \bottomrule
  \end{tabular}
\end{table*}

\section{Proof for the Globally Lipschitz Continuous Proposition}
\label{gen_inst}

 \textbf{The Globally Lipschitz Continuous Proposition}. Let $f$ be a continuous piecewise linear function in the interval $[a, b]$. Then $f$ is globally Lipschitz continuous in $[a, b]$. That is, there exists a constant $K > 0$, for any $x \in [a, b]$ and $y \in [a, b]$,
\begin{equation}\label{eq:51}
|f(x) - f(y)| \leq K*|x-y|.
\end{equation}

Proof.  Let ${x_{0} = a < x_{1} < ... < x_{n} = b}$ be the sequence of $n+1$ vertex points in $[a, b]$ where $\{[x_{i-1} , x_{i}]\}_{i=1}^{n}$ represents the beginning and ending points in the x axis of all the linear segments of the function $f$.

For any two variables $x, y\in [a, b]$, and $x < y$, $\exists \ l, h \in \{1, ..., n\}$ and $l \leq h$, such that $x \in [x_{l-1} , x_{i}]$ and $y \in [x_{h-1} , x_{h}]$. And if $l < h$, we have

\begin{equation}\label{eq:48}
\begin{split}
   |f(x)-f(y)| \leq |f(x)-f(x_{l})| + |f(x_{l})-f(y)| \\
\end{split}
\end{equation}

And furthermore, if $l < h-1$, we have

\begin{equation}\label{eq:481}
\begin{split}
   |f(x)-f(y)|   \leq & |f(x)-f(x_{l})| + |f(x_{l})-f(x_{l+1})| \\
   & + |f(x_{l+1})-f(y)| \\
\end{split}
\end{equation}

And so on, we have
\begin{equation}\label{eq:482}
\begin{split}
   |f(x)-f(y)| \leq & |f(x)-f(x_{l})|  \\
   & + \sum_{k=l}^{h-2} |f(x_{k})-f(x_{k+1})|  \\
   & + |f(x_{h-1})-f(y)|\\
\end{split}
\end{equation}

Notice that each of the pairs of $(x, x_{l})$, $\{(x_{k}, x_{k+1})\}_{k=l}^{h-2}$ and $(x_{h-1}, y)$ is within a linear segment, We therefore have

\begin{equation}\label{eq:49}
\begin{split}
    & |f(x)-f(x_{l})| = |s_{l}|*|x - x_{l}| \\
    & |f(x_{k})-f(x_{k+1})| = |s_{k}|*|x_{k} - x_{k+1}| \\
    & |f(x_{h-1})-f(y)| = |s_{h-1}|*|x_{h-1} - y|,
\end{split}
\end{equation}
where $s_{i}$ denotes the slope of the linear segment in $[x_{i}, x_{i+1}]$.
From (\ref{eq:48}) and (\ref{eq:49}), we can finally have

\begin{equation}\label{eq:50}
\begin{split}
     |f(x)-f(y)| & \leq \sum_{k=l-1}^{h-1}|s_{k}|*|x - y| \\
    & \leq \sum_{k=0}^{n}|s_{k}|*|x - y|.
\end{split}
\end{equation}

The proposition (\ref{eq:51}) is proven.

\section{Experimental Results from the LSR model with Different Hyper Parameter Values}

The LSR models introduce a new hyper parameter $\mu$, as shown in the equations (9) and (12) in the full paper.
We performed some experiments to decide its value.

We tested some values in the range of [0, $10^{-2}$] including $\mu=0$, and compared their results.
Some results are listed in Fig. \ref{fig:wholes2}. They are best for viewing when being scaled up enough.

The average PSNR (in dB), SSIM and the Perceptual Index (PI) values for a few test data sets are also provided for reference. PI is used in the PIRM-SR Challenge and is regarded as being better with a lower value. PSNR and SSIM are evaluated on the luminance channel in YCbCr color space. Table \ref{Experimental Results} lists these values for the LSRGAN model with $\mu$ taking the values of 0, $10^{-7}$ and $10^{-3}$ respectively for the test data sets of Set14, PIRM-SR and BSD100.

First, we find the value of $\mu$ being 0 is less effective in keeping the image structures and details than $\mu > 0$. For example, in the Zebra image, the horizontal stripes are more solid in the front legs when $\mu > 0$ while some stripes become forked with $\mu = 0$; the stripes in the back legs are more faithful to the GT images with $\mu > 0$ than with $\mu = 0$; in the Vegetable image, the number 2.50 on the price tag is more solid when $\mu > 0$; the hat and its decor in the Lenna image are aliased when $\mu = 0$ while
they look nice and sharp in the results when $\mu > 0$, especially the top of hat contains a lot more details when $\mu = 10^{-3}$.

Second, the value of $\mu$ being $10^{-7}$ usually gives sharp details, but they can be excessive sometimes. For example, the whiskers under the nose on the Baboon image and road lines in the country road image are sharper when $\mu = 10^{-7}$. However, some details can be excessive when $\mu = 10^{-7}$. For example, the tree branches on the right side of in the country road image become brush-stroke like and unnatural and the vertical/tilted lines on the hat of the Lenna image seem sharpened too much in the results with $\mu$ being $10^{-7}$. With the value of $10^{-3}$, the results look more balanced overall.

It is recognized in the research community that numerical scores such as PSNR, SSIM and PI alone are not suitable for differentiating and evaluating image perceptual quality because they do not always correlate very well with subjective quality \cite{DBLP:journals/corr/abs-1809-07517,ledig2017photo}, but they can be still important references. By checking the average PSNR, SSIM and PI values in Table \ref{Experimental Results}, we find that, in contrast to the subjective quality difference we have just checked above, the numerical scores are only marginally different. Nevertheless, the value of $\mu$  being $10^{-3}$ gives 9 best scores out of the 12 comparisons. Subjective quality is also generally better with $\mu$  being $10^{-3}$ although it is not always the case.

The value of $\mu$  being $10^{-3}$ is therefore used in the experiments reported in the paper.

The section of Experiments of the paper thoroughly evaluates how the LSR GAN models help make the internal image structures and details better kept than the models without the LSR.

\end{document}